\documentclass[aps,prd,twocolumn,nofootinbib]{revtex4}

\usepackage{graphicx}

\begin{document}
\title{A Trick to Improve the Efficiency of Generating Unweighted $B_c$ Events from BCVEGPY}

\author{Xian-You Wang$^{1}$}
\email[email: ]{xianyou.wang@cern.ch}

\author{Xing-Gang Wu$^{1,2}$}
\email[email: ]{wuxg@cqu.edu.cn}

\address{$^{1}$ Department of Physics, Chongqing University, Chongqing 401331, P.R. China\\
$^{2}$ SLAC National Accelerator Laboratory, 2575 Sand Hill Road, Menlo Park, CA 94025, USA}

\date{\today}

\begin{abstract}
In the present paper, we provide an addendum to improve the efficiency of generating unweighted events within PYTHIA environment for the generator BCVEGPY2.1 [C.H. Chang, J.X. Wang and X.G. Wu, Comput.Phys.Commun.{\bf 174}, 241(2006)]. This trick is helpful for experimental simulation. Moreover, the BCVEGPY output has also been improved, i.e. one Les Houches Event common block has been added so as to generate a standard Les Houches Event file that contains the information of the generated $B_c$ meson and the accompanying partons, which can be more conveniently used for further simulation. \\



\end{abstract}

\maketitle

\noindent{\bf NEW VERSION PROGRAM SUMMARY} \\

\noindent{\it Title of program} : BCVEGPY2.1a \\

\noindent{\it Program obtained from} : CPC Program Library \\

\noindent{\it Reference to original program} : BCVEGPY2.1 \\

\noindent{\it Reference in CPC} : Comput. Phys. Commun. {\bf 175},
624(2006) \\

\noindent{\it Does the new version supersede the old program}: No\\

\noindent{\it Computer} : Any LINUX based on PC with FORTRAN 77 or FORTRAN 90 and GNU C compiler as well.\\

\noindent{\it Operating systems} : LINUX \\

\noindent{\it Programming language used} : FORTRAN 77/90 \\

\noindent{\it Memory required to execute with typical data} : About 2.0 MB \\

\noindent{\it No. of bytes in distributed program, (including PYTHIA6.4)} : About 1.5 MB.\\

\noindent{\it Distribution format} : .tar.gz \\

\noindent{\it Nature of physical problem} : Hadronic Production of $B_c$ meson and its excited states. \\

\noindent{\it Method of solution} : To generate weighted and unweighted $B_c$ events within PYTHIA environment effectively. \\

\noindent{\it Restrictions on the complexity of the problem} : Hadronic production of ($c\bar{b}$)-quarkonium via the gluon-gluon fusion mechanism are given by the `complete calculation approach'. The simulation of $B_c$ events is done within PYTHIA environment. \\

\noindent{\it Reasons for new version} : More and more data are accumulated at the large hadronic collider, it would be possible to make precise studies on $B_c$ meson properties, such as its lifetime, mass spectrum and etc.. The BCVEGPY has been adopted by several experimental groups due to its high efficiency in comparison to that of PYTHIA. However, to generate unweighted events with PYTHIA inner mechanism as programmed by the previous verison is still time-consuming. So it would be helpful to improve the efficiency for generating unweighted events within PYTHIA. Moreover, it would be better to use an uniform and standard output format for further detector simulation.  \\

\noindent{\it Typical running time} : Typical running time is machine and user-parameters dependent. I) To generate $10^6$ weighted $S$-wave ($c\bar{b}$)-quarkonium events (IDWTUP=3), it will take about 40 minutes on a 1.8 GHz Intel P4-processor machine. II) To generate unweighted $S$-wave ($c\bar{b}$)-quarkonium events with PYTHIA inner structure (IDWTUP=1), it will take about 20 hour on a 1.8 GHz Intel P4-processor machine to generate 1000 events. III) To generate $10^6$ unweighted $S$-wave ($c\bar{b}$)-quarkonium events with the present trick (IDWTUP=1), it will take 17 hour on a 3.16 GHz Intel E8500 processor machine. Moreover, it can be found that the running time for the $P$-wave ($c \bar{b}$)-quarkonium production is about two times longer than the case of S-wave production under the same conditions. \\


\noindent{\it Summary of the changes} :  \\

\noindent (1) The generator BCVEGPY \cite{bcvegpy1,bcvegpy2,bcvegpy21} has been programmed to generate $B_c$ events under PYTHIA environment \cite{pythia}, which has been frequently adopted for theoretical and experimental studies, e.g. Refs. \cite{the1,the2,the3,the4,the5,the6,the7,exp0,exp1,exp2,exp3,exp4,exp5,exp6}. It is found that each experimental group shall have its own simulation software architecture, and the users will spend a lot of time to write an interface so as to implement BCVEGPY into their own software. So it would be better to supply a standard output.

The LHE format becomes a standard format \cite{lhe}, which is proposed to store process and event information from the matrix-element-based generators. The users can pass these parton-level information to the general event generators like PYTHIA and HERWIG \cite{herwig} for further simulation. For such purpose, we add two common blocks in genevent.F. One common block is called as bcvegpy\_pyupin and the other one is write\_lhe. bcvegpy\_pyupin, which is similar to PYUPIN subroutine in PYTHIA, stores the initialization information in the HEPRUP common block.\\

\begin{widetext}
\noindent INTEGER MAXPUP\\
PARAMETER (MAXPUP=100)\\
INTEGER IDBMUP,PDFGUP,PDFSUP,IDWTUP,NPRUP,LPRUP\\
DOUBLE PRECISION EBMUP,XSECUP,XERRUP,XMAXUP\\
COMMON/HEPRUP/IDBMUP(2),EBMUP(2),PDFGUP(2),PDFSUP(2),\\
\&IDWTUP,NPRUP,XSECUP(MAXPUP),XERRUP(MAXPUP),\\
\&XMAXUP(MAXPUP),LPRUP(MAXPUP)\\
\end{widetext}

The write\_lhe, which is similar to PYUPEV subroutine in pythia, stores the information of each separate event in the HEPEUP common block.\\

\begin{widetext}
\noindent INTEGER MAXNUP\\
PARAMETER (MAXNUP=500)\\
INTEGER NUP,IDPRUP,IDUP,ISTUP,MOTHUP,ICOLUP\\
DOUBLE PRECISION XWGTUP,SCALUP,AQEDUP,AQCDUP,PUP,VTIMUP,\\
\&SPINUP\\
COMMON/HEPEUP/NUP,IDPRUP,XWGTUP,SCALUP,AQEDUP,AQCDUP,\\
\&IDUP(MAXNUP),ISTUP(MAXNUP),MOTHUP(2,MAXNUP),\\
\&ICOLUP(2,MAXNUP),PUP(5,MAXNUP),VTIMUP(MAXNUP),\\
\&SPINUP(MAXNUP)\\
\end{widetext}

The above information can be separately exported to two files `bcvegpy.init' and `bcvegpy.evnt'. Then with the help of the PYTHIA subroutine PYLHEF, we can combine these two files into a single LHE file `bcvegpy.lhe'.\\

\noindent (2) As for previous version of BCVEGPY \cite{bcvegpy1,bcvegpy2,bcvegpy21}, its main concern is to improve theoretical estimations, i.e. to obtain a precise result as fast as possible. For the purpose, the VEGAS \cite{vegas} running together with IDWTUP=3 is usually appreciated. Firstly, by running VEGAS, one can obtain a sampling importance function, and then the phase-space points are generated according to the relative importance of their corresponding differential cross-sections. This sampling importance function is useful, since the integrand for the phase space integration can be regularized for obtaining a high precision total/differential cross section \cite{vegas}. Secondly, by setting IDWTUP=3, all parton-level events when input to PYTHIA shall always be accepted, i.e. almost all the calculated points are effective ones and shall be evolved as final events. As a combination of these two choices, one can generate events within the so-called importance sampling approach, i.e. the parton-level events have been effectively input into PYTHIA with proper weighted phase space. This is one of the reason why BCVEGPY runs so fast \footnote{Nevertheless, the dominate reason for its quick running is due to the use of improved helicity amplitude approach to deal with the hard scattering amplitude as described in detail in Ref.\cite{bcvegpy1}. }.

Experimentally, it is the unweighted inputs other than the weighted inputs that are usually needed. By default BCVEGPY generates unweighted events following the same scheme of PYTHIA.  For the purpose, one have to set PYTHIA parameter IDWTUP=1, then all events in a run are treated on an equal footing. And the parton-level events are uniformly generated within the whole phase space, but shall with a weight (XWGTUP) when input to PYTHIA, and then these events are accepted or rejected with probability XWGTUP/XMAXUP (XMAXUP is the maximum weight of the process). As most of the calculated parton-level events are rejected by this criteria, such a naive scheme is usually time-consuming. For example, by setting IDWTUP=1, it takes about 20 hours on a 1.8 GHz Intel P4-processor machine to generate 1000 $B_c$ events. It is too slow if one wants to simulate large amount of events.

So, we present a new scheme to improve the efficiency of generating unweighted events within PYTHIA environment. For the purpose, we first unweight the events by using the standard hit-and-miss technique \cite{unweight}. Then, we set IDWTUP=1, but we replace XMAXUP to be a reference weight that can be generated by earlier exploratory run \footnote{The function of the reference weight is similar to that of XMAXUP, and for convenience, we still use the label XMAXUP to store its value. }. Each parton-level event shall result in a ratio between the weight XWGTUP to the reference weight. When such ratio is smaller than one, the event shall be retained or rejected based on whether the ratio is greater or less than the return value of random number generating function PYR(0). When the ratio is larger than one, its integral part stands for the basic number the same events to be generated. And the basic number shall be increased by one or unchanged based on whether its decimal part is greater or less than a random number PYR(0). It is found that a larger reference weight means more events to be rejected and less efficiency. So, to set a proper reference weight is a tricky problem, which requires exploratory running under one's own experimental cuts and should has negligible derivation from theoretical estimations. For example, one may observe that for ATLAS transverse momentum cut $P_{Tcut}=4$ GeV and the rapidity cut $y_{cut}=2.5$, only few events' weight are larger than the value of $10^6$ pb. So we can set this valve as the reference weight, which is reasonable for the LHC $B_c$ production simulation.

\begin{figure}
\centering
\includegraphics[width=0.40\textwidth]{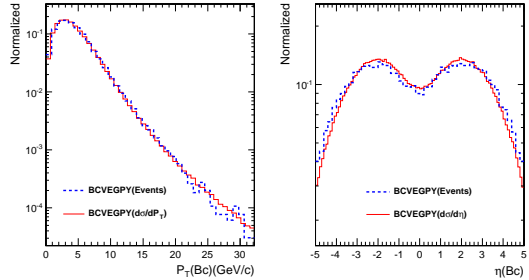}
\caption{Comparison of the normalized transverse momentum ($P_{T}$) and pseudo-rapidity ($\eta$) distributions between the weighted ones (IDWTUP=3) and the un-weighted ones (IDWTUP=1), which are represented by solid line and dotted lines respectively. }
\label{fig1}
\end{figure}

The code for the newly suggested unweighted scheme is put in the main program `bcvegpy.F'. As a cross check, we present the weighted and unweighted $B_c$ transverse momentum $P_T$-distribution and pseudo-rapidity $\eta$-distribution in Fig.(\ref{fig1}), which are represented by solid line and dotted lines respectively. The unweighted distributions are obtained by setting IDWTUP=1 and using the new strategy described above, while the weighted distributions are obtained by setting IDWTUP=3. It can be found that the weighted and unweighted distributions are coincide with each other. Since the new unweighted scheme is time-saving in comparison with that of the previous one, so it can be adopted for generating large amount of events, which may suit the needs of the newly accumulating data at LHC.

\begin{figure}
\centering
\includegraphics[width=0.40\textwidth]{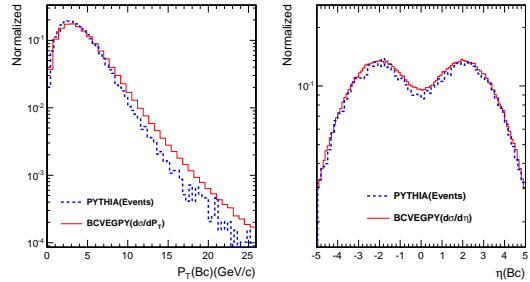}
\caption{Comparison of the normalized transverse momentum ($P_{T}$) and pseudo-rapidity ($\eta$) distributions derived by BCVEGPY and PYTHIA, which are represented by red-solid line and blue-dotted lines respectively.}
\label{fig2}
\end{figure}

Moreover, to compare and validate the result of BCVEGPY, we also use PYTHIA's inner mechanism to generate $B_c$ meson at the very beginning. For such purpose, we set MSEL=5 to generate $b\bar{b}$ first, and then add one filter (PDGID=541) that searches and stores every event that shall result in at least one $B_c$ meson. Next, we use the external decay generator EVTGEN \cite{evtgen}, which can be running under the PYTHIA environment, to simulate the $B_c$ decay channel $B_c\to J/\psi+\pi$. The remaining parameters for PYTHIA are taken to be their default values. Moreover, in doing the calculation, we set the collision energy of LHC to be the present value of $7$ TeV. The comparison is presented in Fig.(\ref{fig2}), which shows that the normalized distributions for PYTHIA and BCVEGPY agree with each other well. Such an observation also agrees with the previous conclusion drawn by Ref.\cite{exp0}. \\

{\bf Acknowledgments}: We thank Prof. C.H. Chang, Prof. Y.N. Gao, Dr. J.B. He, Dr. Z.W. Yang for helpful discussions. We thank Dr. C. Philips and Dr. Z.C. Tang for testing the program and providing helpful suggestions to solve the compiling problems within different running environment. This work was supported in part by the Fundamental Research Funds for the Central Universities under Grant No.CDJXS1102209 and the Program for New Century Excellent Talents in University under Grant No. NCET-10-0882, and by Natural Science Foundation of China under Grant No.10805082 and No.11075225.

\end{document}